\documentclass[twoside,twocolumn,9pt]{article}
\usepackage{extsizes}
\usepackage[super,sort&compress,comma]{natbib} 
\usepackage[version=3]{mhchem}
\usepackage[left=1.5cm, right=1.5cm, top=1.785cm, bottom=2.0cm]{geometry}
\usepackage{balance}
\usepackage{mathptmx}
\usepackage{sectsty}
\usepackage{graphicx} 
\usepackage{lastpage}
\usepackage[format=plain,justification=justified,singlelinecheck=false,font={stretch=1.125,small,sf},labelfont=bf,labelsep=space]{caption}
\usepackage{float}
\usepackage{fancyhdr}
\usepackage{fnpos}
\usepackage[english]{babel}
\addto{\captionsenglish}{%
  
}
\usepackage{array}
\usepackage{droidsans}
\usepackage{charter}
\usepackage[T1]{fontenc}
\usepackage[usenames,dvipsnames]{xcolor}
\usepackage{setspace}
\usepackage[compact]{titlesec}
\usepackage{hyperref}
\usepackage{epstopdf}%This line makes .eps figures into .pdf - please comment out if not required.

\definecolor{cream}{RGB}{222,217,201}

\begin{document}

\pagestyle{fancy}
\thispagestyle{plain}
\fancypagestyle{plain}{
%%%HEADER%%%
\renewcommand{\headrulewidth}{0pt}
}
%%%END OF HEADER%%%

%%%PAGE SETUP - Please do not change any commands within this section%%%
\makeFNbottom
\makeatletter
\renewcommand\LARGE{\@setfontsize\LARGE{15pt}{17}}
\renewcommand\Large{\@setfontsize\Large{12pt}{14}}
\renewcommand\large{\@setfontsize\large{10pt}{12}}
\renewcommand\footnotesize{\@setfontsize\footnotesize{7pt}{10}}
\makeatother

\renewcommand{\thefootnote}{\fnsymbol{footnote}}
\renewcommand\footnoterule{\vspace*{1pt}% 
\color{cream}\hrule width 3.5in height 0.4pt \color{black}\vspace*{5pt}} 
\setcounter{secnumdepth}{5}

\makeatletter 
\renewcommand\@biblabel[1]{#1}            
\renewcommand\@makefntext[1]% 
{\noindent\makebox[0pt][r]{\@thefnmark\,}#1}
\makeatother 
\renewcommand{\figurename}{\small{Fig.}~}
\sectionfont{\sffamily\Large}
\subsectionfont{\normalsize}
\subsubsectionfont{\bf}
\setstretch{1.125} %In particular, please do not alter this line.
\setlength{\skip\footins}{0.8cm}
\setlength{\footnotesep}{0.25cm}
\setlength{\jot}{10pt}
\titlespacing*{\section}{0pt}{4pt}{4pt}
\titlespacing*{\subsection}{0pt}{15pt}{1pt}
%%%END OF PAGE SETUP%%%

%%%FOOTER%%%
\fancyfoot{}
\fancyfoot[LO,RE]{\vspace{-7.1pt}\includegraphics[height=9pt]{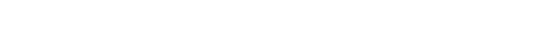}}
\fancyfoot[CO]{\vspace{-7.1pt}\hspace{13.2cm}\includegraphics{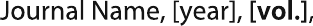}}
\fancyfoot[CE]{\vspace{-7.2pt}\hspace{-14.2cm}\includegraphics{head_foot/RF}}
\fancyfoot[RO]{\footnotesize{\sffamily{1--\pageref{LastPage} ~\textbar  \hspace{2pt}\thepage}}}
\fancyfoot[LE]{\footnotesize{\sffamily{\thepage~\textbar\hspace{3.45cm} 1--\pageref{LastPage}}}}
\fancyhead{}
\renewcommand{\headrulewidth}{0pt} 
\renewcommand{\footrulewidth}{0pt}
\setlength{\arrayrulewidth}{1pt}
\setlength{\columnsep}{6.5mm}
\setlength\bibsep{1pt}
%%%END OF FOOTER%%%

%%%FIGURE SETUP - please do not change any commands within this section%%%
\makeatletter 
\newlength{\figrulesep} 
\setlength{\figrulesep}{0.5\textfloatsep} 

\newcommand{\topfigrule}{\vspace*{-1pt}% 
\noindent{\color{cream}\rule[-\figrulesep]{\columnwidth}{1.5pt}} }

\newcommand{\botfigrule}{\vspace*{-2pt}% 
\noindent{\color{cream}\rule[\figrulesep]{\columnwidth}{1.5pt}} }

\newcommand{\dblfigrule}{\vspace*{-1pt}% 
\noindent{\color{cream}\rule[-\figrulesep]{\textwidth}{1.5pt}} }

\makeatother
%%%END OF FIGURE SETUP%%%

%%%TITLE, AUTHORS AND ABSTRACT%%%
\twocolumn[
  \begin{@twocolumnfalse}
{\includegraphics[height=30pt]{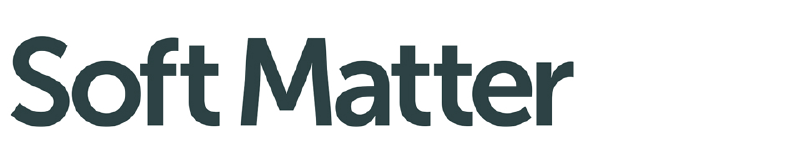}\hfill\raisebox{0pt}[0pt][0pt]{\includegraphics[height=55pt]{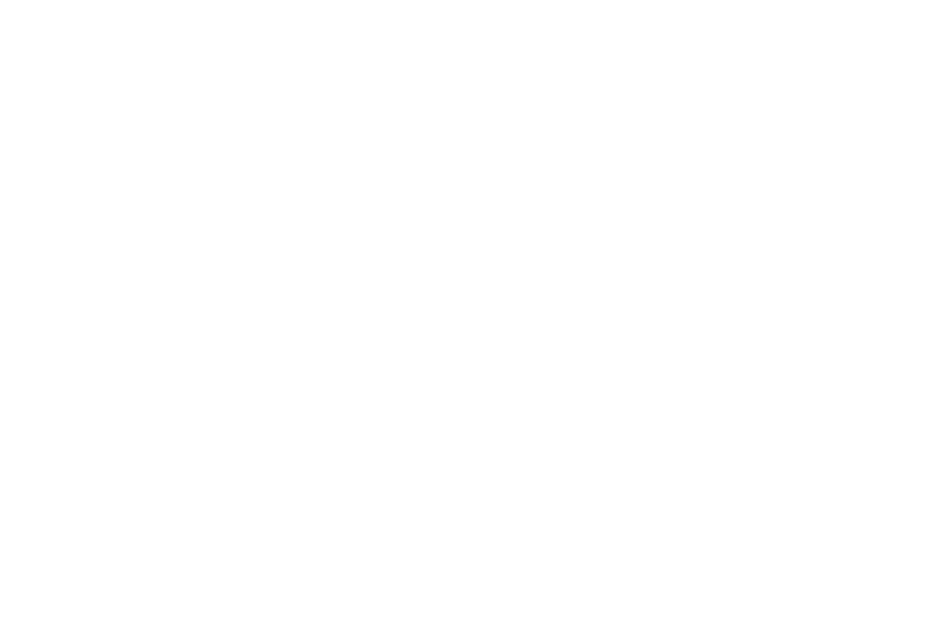}}\\[1ex]
\includegraphics[width=18.5cm]{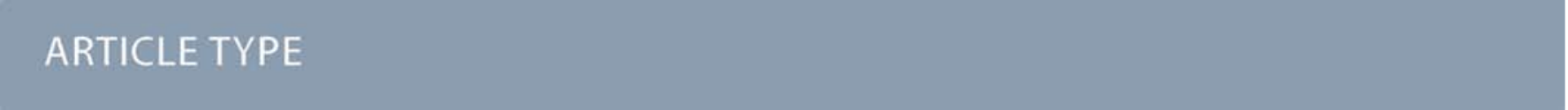}}\par
\vspace{1em}
\sffamily
\begin{tabular}{m{4.5cm} p{13.5cm} }

%%%%%%%%%%% This is where we input our names affiliations etc. %%%%%%%%%%%%%

\includegraphics{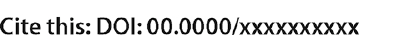} & \noindent\LARGE{\textbf{Braiding dynamics in 
semiflexible filament bundles under oscillatory forcing}} \\
\vspace{0.3cm} & \vspace{0.3cm} \\

 & \noindent\large{ Valentin M.~Slepukhin\textit{$^{a}$} and 
 Alex J.~Levine\textit{$^{a,b}$}} \\

\includegraphics{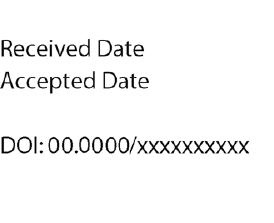} & \noindent\normalsize{We examine the nonequilibrium 
production of topological defects -- braids -- in semiflexible 
filament bundles under cycles of compression and tension.  During these cycles, the period of 
compression facilitates the thermally activated pair production of braid/anti-braid pairs, which then may 
separate when the bundle is under tension.  As result, 
appropriately tuned alternating periods of compression and extension should lead to the 
proliferation of braid defects in a bundle so that linear density of these pairs far exceeds that 
expected in thermal equilibrium. Secondly, we examine the slow extension of braided bundles under 
tension, showing that their end-to-end length creeps nonmonotonically under a fixed force due to 
braid deformation and the motion of the braid pair along the bundle.  
We conclude with a few speculations regarding experiments on semiflexible filament bundles and their networks.}

\end{tabular}

\end{@twocolumnfalse} \vspace{0.6cm}

  ]
%%%END OF TITLE, AUTHORS AND ABSTRACT%%%

%%%FONT SETUP - please do not change any commands within this section
\renewcommand*\rmdefault{bch}\normalfont\upshape
\rmfamily
\section*{}
\vspace{-1cm}

%%%FOOTNOTES%%%

\footnotetext{\textit{$^{a}$~Department of Physics \& Astronomy, University of California, Los Angeles. 90095  USA. }}
\footnotetext{\textit{$^{b}$~Department of Chemistry \& Biochemistry, University of California, Los Angeles. 90095  USA. }}

%Please use \dag to cite the ESI in the main text of the article.
%If you article does not have ESI please remove the the \dag symbol from the title and the footnotetext below.
%\footnotetext{\dag~Electronic Supplementary Information (ESI) available: [details of any supplementary information available should be included here]. See DOI: 00.0000/00000000.}
%additional addresses can be cited as above using the lower-case letters, c, d, e... If all authors are from the same address, no letter is required

%%%END OF FOOTNOTES%%%

%%%%%%%%%%%%%%%%%%%%%% SECTION: INTRODUCTION%%%%%%%%%%%%%%%%%%%%%%%

\section{Introduction}

The mechanics of semiflexible filaments has been a subject of broad interest both for its 
role in the mechanics of the cytoskeleton and as a testing ground for various 
principles of polymer and soft condensed matter physics.  One 
feature of biological filaments, such as F-actin and collagen (a key constituent of the 
extracellular matrix) is their
ability to form densely cross-linked bundles.  These bundles are 
composed of a number of nearly parallel filaments 
cross linked by one of a variety of specialized proteins. 

Previous research has focused on the collective mechanical response of permanently cross-linked 
filament bundles~\cite{Heussinger2007},
showing that the bundle inherits a complex, scale-dependent bending modulus due to cross linking, even though 
the bending mechanics of the constituent filaments is comparatively simple. In most circumstances of 
biological interest, cross linkers detach and attach to 
filament bundles and their networks.  As a result, such structures acquire a viscoelastic response -- their 
stress relaxation has a complex time dependence and these systems dissipate work not only 
through viscous dissipation
in the surrounding fluid but also by linker unbinding.   As a result, the collective 
mechanical response of networks of filament bundles have a nontrivial low-frequency viscoelastic response at 
frequencies below a characteristic linker unbinding rate~\cite{Weitz2010,Mueller2014}. 

We explore here particular type of stress relaxation through the production and movement of defects in 
cross-linked bundles. Previously, we have shown that bundles support a set of 
topological defects -- loop, braids, and dislocations~\cite{slepukhin2021topological}. 
The lifetime of these defects is quite long, growing with the length of the bundle, since they cannot be removed 
by local rearrangements of the cross linking on the bundle.  Defects, however, can be 
produced in defect/anti-defect pairs by local rearrangements, and defect pair production is 
predicted to be enhanced by applied compressive loads~\cite{slepukhin2021thermal}.  In this article, we report on 
theoretical studies of defect pair production under reciprocal mechanical deformations and consider 
how the proliferation of defects affects 
the force-extension relation of a bundle in a pulling-velocity-dependent manner.  

The motion of topological 
defects plays a critical role in the long-time plastic deformation of crystalline solids under mechanical loading.
Defect motion in plays a similar role in the slow relaxation of bundles under load.  Mechanical loading can also
generate defect pair production.  We first consider pair production in cycles of compression and extension of one
bundle.  We then examine the force extension relation of defected bundles by examining the extension of the 
bundle as a function of time for fixed force. 
%%%%% FIGURE: setup %%%%%
\begin{figure*}[htpb]
  \centering
  \includegraphics[width=\linewidth]{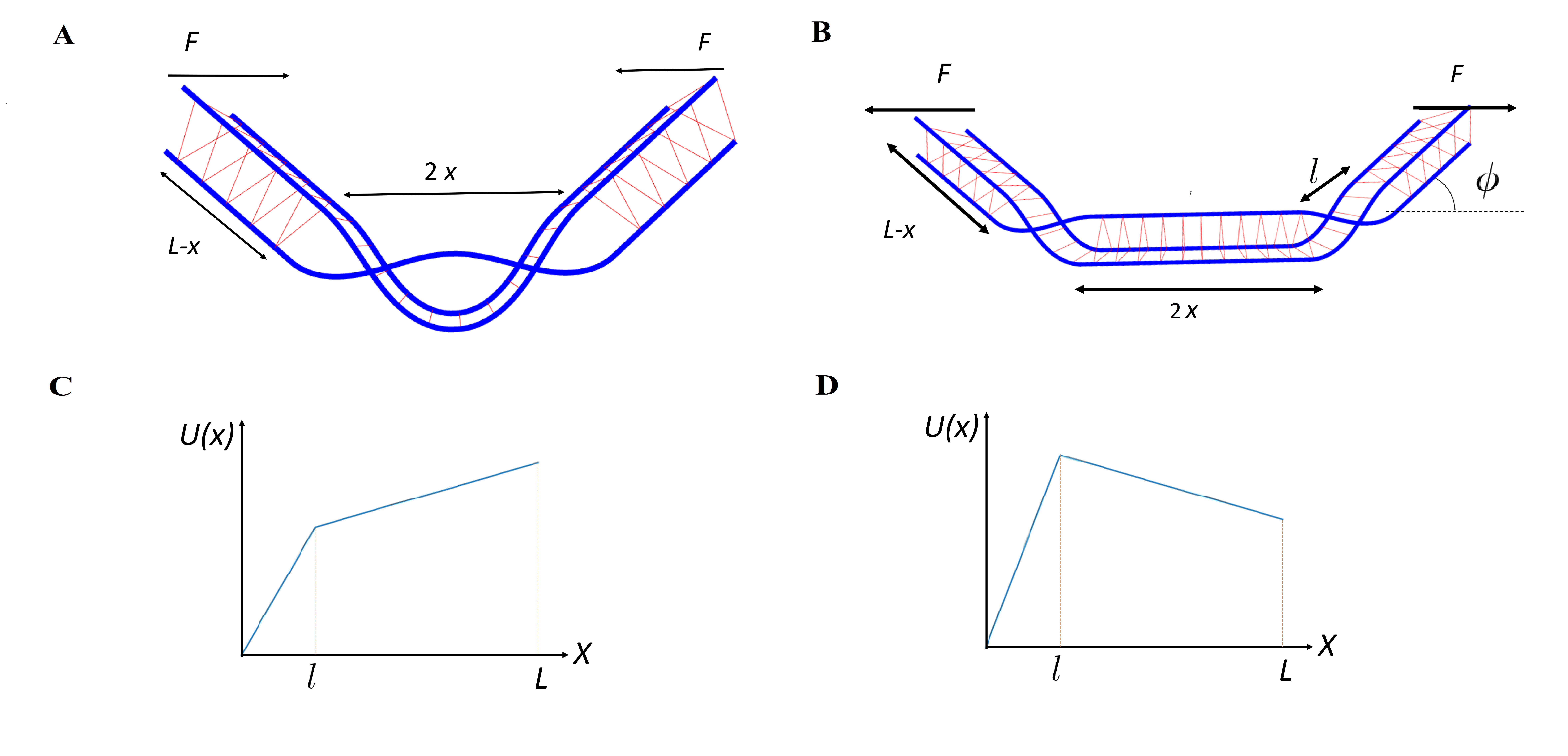}
\caption{Schematic illustration: (A) A pair of braids produced under compression force $F$ 
on a semiflexible filament 
bundle of length $2L$ (and comprised of three filaments), but not separated. $2 x$ 
is the size of the defected region. The filaments are shown in blue while the cross links are shown in red. 
(B) Pair of braids is separated under tension. The
braids produce a kink with angle $\phi$. $2 x$ is the distance 
between braids (including their size). In the lower figures we show the piece-wise linear 
potential $U(x)$ under compression 
(C) and under  extension (D) . The left part of the potential $x < l$ corresponds to the production of 
defects, the while right $x > l$ controls the separation of the defects. }
  \label{fig:setup}
\end{figure*}
%%%%%%%%%%%%%%%%%%%%%%%%%%%%%%%%%%%%%%%%%%%%%%%%%%%%%%%%%%

\section{The piece-wise linear defect potential and the defect distribution}

We have shown that the production of braid/anti-braid pairs in a compression bundle can be mapped onto the 
Kramers escape problem in one dimension using a single reaction 
coordinate~\cite{kramers1940brownian, slepukhin2021thermal}.  The two defects must be 
first produced together in the form of defect/anti-defect pairs, which requires energy $E_{\rm defect}$.  The formation of one defect introduces a 
length mismatch between the filaments involved that is then compensated by the second, anti-defect.  As a result,
defect pair production entails only local rearrangements of cross linkers.  During this 
point in the thermally-activated
production of the defect pair, we may take the single reaction coordinate to represent the total length 
``exchanged'' between defects.  Once the defects separate so that a region of cross-linked bundle appears between
them, the defects can no longer exchange length, but they may separate along the bundle by reptative motion. 
Most importantly, since the defects generate localized bends or {\em kinks} under external loading, the motion of
the defects changes the end-to-end distance of the bundle under load.  By separating the energy of the system 
decreases.  During this separation (under external loading) the distance between the 
defects plays the role of the reaction coordinate.

Taking these two aspects of the problem together, we may consider the stochastic pair production process as the 
thermal escape of a single fictitious particle, representing the reaction coordinate 
$x$, in an approximately piece-wise linear 
potential.  Before defect pair separation $x<l$, where $l$ is the size of the defect at the moment 
of the separation, the potential increases linearly as more length is exchanged 
between the defect pair. The exact answer would require taking into account all different pathways 
from properly cross-linked bundle to the bundle with braids. We make the simplest approximation, i.e., a 
linear potential, motivated by the fact that we need to remove number of cross links proportional to the 
size of the uncross-linked region. Then the effective potential is $U(x) = A x$, with $A = E_{\rm defect} / l$. 
The energy of the defect incorporates bending energy of the filaments, missing binding energy of the 
cross links not present in the defected region, and the work of the applied force: 
\begin{equation}
    E_{\rm defect} = E_{\rm bending} (\phi) + E_{\rm binding} (\phi) - 2 F (L - x) \cos(\phi) - 2 F x,
\end{equation}
where $\phi$ is the angle formed by a single braid (see Figure~\ref{fig:setup} ), $F$ is a 
force acting on the ends of the bundle (positive sign is chosen for the extension), 
$2 L$ is a length of the bundle. The angle $\phi$ is 
determined by energy minimization with respect to it. As defects separate, $x$ grows. The angle also 
changes, but as soon as $x \ll L$, this change contributes to the energy at the next order in the 
small parameter $x/L$. 
We address the explicit dependence of the angle on coordinate in the Section~\ref{sec:dynamic}. 
Omitting this effect, we can again write an effective potential $U(x) = A l + B (x - l) $ 
with $B  = 2 F (\cos \phi - 1)$.
Thus, we may explore the pair production process using the potential 
\begin{equation}
\label{eq:potential}
    U(x) = 
    \begin{cases}
    A x , x < l 
    \\
    A l + B (x - l) , l < x < L
    \\
    \infty , x > L
    \end{cases}
\end{equation}
Before considering dynamics, we use this potential to consider the equilibrium distribution of defects on 
a bundle of length $L$.  Specifically, we consider the equilibrium separation of two defects.  
Using Eq.~\ref{eq:potential} it is trivial to write the probability distribution in this potential: $p(x) = \frac{1}{Z} e^{-\beta U(x)}$ where the partition 
function $Z$ is given by 
\begin{equation}
\label{eq:partition-function}
    Z = \int_0^L  e^{-\beta U(x)} dx,
\end{equation}
and $\beta = 1/k_{\rm B}T$.  
A straightforward calculation arrives at the partition function written as the sum of 
two parts corresponding to the two pieces of the potential
\begin{equation}
    Z = Z_1 + Z_2.
\end{equation}
with 
\begin{equation}
    Z_1 = \frac{1}{\beta A} (1  -   e^{-\beta A l} )
\end{equation}
and
\begin{equation}
    Z_2 =    \frac{1}{\beta B}  e^{ -\beta (A - B) l}   ( e^{ -\beta B l } - e^{ -\beta B L }  ).
\end{equation}
Taking the ratio of these partition sums we obtain the 
ratio of observable separated braid pairs to strongly interacting and co-localized braids:
\begin{equation}
    \label{eq:equilibrium-ratio}
     Z_2/ Z_1 = \frac{ A e^{ -\beta (A - B) l}   ( e^{ -\beta B l } - e^{ -\beta B L }  )}{  B (1  -   e^{-\beta A l} )}
\end{equation}
From the above we see that in thermal equilibrium we expect there to be a low density of 
separated braids, at least at low temperatures ($T \approx 300$K). 
Specially, if the thermal energy is much lower than the cross-linker binding energy 
(which is typically true in biopolymer filament systems) we expect an exponentially small 
density of braids ($\propto e^{-E_{\rm braid} \beta} $, where for 
known filaments $\beta E_{\rm braid} \gg 1$, the smallest is for DNA with $\beta E_{\rm braid} 
\approx 50$. Braid pairs, however, can be generated either during bundle formation or via cycles 
of compression and expansion, as would be expected in a bundle network under reciprocal shear.

\section{The nonequilibrium braid distribution in a time-dependent potential}

When one applies a time-varying force, the effective potential controlling the 
production and motion of the braids also changes in time.  As a result, we cannot 
rely on the equilibrium distribution discussed in the previous section. Instead, 
we have to solve the Smoluchowski diffusion equation for defect density $\rho(t,x)$:
\begin{equation}
\label{eq:smoluchowski}
    \partial_t \rho (t,x) = D \partial_x \left[ \left(\partial_x - \beta F(t,x) \right) \rho(t,x)\right],
\end{equation}
where the force now takes the form
\begin{equation}
    F(t, x) = 
    \begin{cases}
    A(t)  , x < l 
    \\
    B(t) , l < x < L.
    \end{cases}
\end{equation}
During compression $A(t) = A_c$ and  $B(t) = B_c$. During expansion $A(t) = A_s$, $B(t) = B_s$ 
(note that $B_s$ is negative). In the above, $D$ is the defect diffusion constant.  
We cannot solve this equation analytically, however, we can provide a qualitative analysis. 
To simplify, we assume that $A$ and $B$ are fixed during each period of compression and expansion. 
We explore how the defect production rate depends on the lengths of these periods of compression and 
extension. We also estimate the maximal defect production rate. 

The transport time from $0$ to $l$ in the potential is controlled by the 
constant $A$.  This is braid pair production
rate when the braids have stored length of $l$.  This problem is simply 
first passage time~\cite{kramers1940brownian} 
to reach $l$ in the linearly increasing potential, which we may estimate to be
\begin{equation}
    T_{0 l} = \frac{1}{D} \frac{e^{\beta A  l}}{\beta^2 A^2 }.
\end{equation}
Similarly, we estimate the transport time from $L$ to $l$.  This gives an estimate of the lifetime of
the braid pair, since when their separation returns to $l$ they will likely annihilate.  
Here we must distinguish two limiting cases. In the first case, we consider purely 
diffusive braid motion and in the
second, we look at the deterministic transport of the braids under an applied force using a mobility 
set by the diffusion 
constant and the Einstein relation.  We find
\begin{equation}
    T_{L l} = 
    \begin{cases}
    L^2 / D , B \beta > 1 / L 
    \\
    L / (B \beta D) , B \beta < 1 /L.
    \end{cases}
\end{equation}
In general, where we expect there to be both diffusive and advective defect motion, we find that the
time for defects to recombine is 
\begin{equation}
    T_{L l} = 
    \frac{L^2}{D (1 + \beta B L)  }
\end{equation}
In the limit of large $L$,  $T_{L l} \gg T_{0 l}$ so that the time for distant defect pairs to come 
together and potentially annihilate is much greater than their production time.  If braids are able to separate 
sufficiently, we expect this ordering of time scales to be valid and thus predict 
braid proliferation on the bundle. 
%%%%% FIGURE: diffusion %%%%%
\begin{figure}[htpb]
  \centering
  \includegraphics[width=\linewidth]{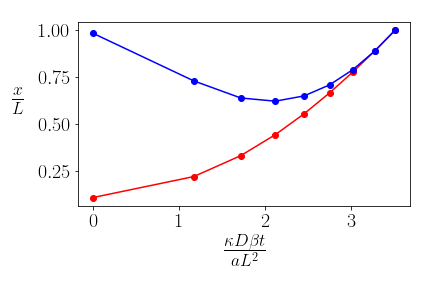}
\caption{The time evolution of the distance between the defects (red) and the distance between the 
ends of the bundle (blue) under a constant applied tensile force $F = 0.9 \frac{\kappa}{a L}$ plotted as a 
function of non-dimensionalized time.  The cross linker binding energy is $\mu = 4 \frac{\kappa}{a^2}$, 
where $\kappa$ is bending modulus of a single filament, and $a$ is the size of the cross-link plus 
two radii of the filament (the distance between the center lines of the cross linked filaments).}
  \label{fig:diffusion}
\end{figure}
%%%%%%%%%%%%%%%%%%%%%%%%%%%%%%%%%%%%%%%%%%%%%%%%%%%%%%%%%%

Because the production time $T_{0 l}$ has an exponential dependence on $A$, 
\begin{equation}
    T_{0 l}^{\rm compression} \ll    T_{0 l}^{\rm stretching}
\end{equation}
since the $A$ parameter is much larger under stretching than it is under compression: $A_c < A_s$.
If we choose the time-dependence of the applied force so that 
the stretching time $\tau_s$ and compression $\tau_c$ satisfy the inequalities
\begin{equation}
    T_{0 l}^{\rm compression} \ll \tau_c \ll \tau_s \ll T_{0 l}^{\rm stretching} \ll T_{L l},
\end{equation}
we may analyse the dynamics of the system using a few approximations.  

Since the braid production rate during compression is small compared to the compression time: 
$T_{0 l}^{\rm compression} \ll \tau_c$, then during the compression period, the density 
on the left $0<x<l$ equilibrates. Since $\tau_s \ll T_{0 l}^{\rm compression} \ll T_{L l}$,
during the stretching period, the probability density $\rho(x)$ decreases near the potential maximum at $x=l$, 
but it is highly unlikely that thermally excited hopping over the barrier at $l$ occurs. 
Since $\tau_c~\ll~\tau_s~\ll~T_{0 l}^{\rm compression}~\ll~T_{L l}$, the applied force 
is changing sufficiently fast that the density of the right of of the potential $x>l$ 
may be replaced by its time-averaged value.  Moreover, since $\tau_c \ll \tau_s$, 
density on the right of the potential 
is effectively determined by the dynamics during the stretching period. Because the compression 
period is shorter, the 
already produced braids are unlikely to be driven together and annihilate.  
As a result, we conclude that the density distribution on the left is effectively determined
by the compression period and the density on the right by the stretching period. 
Finally, we note that the Smoluchowski diffusion equation requires the
continuity of both the probability density and its current at the boundary $x=l$.  From these 
conditions, we obtain a value of the averaged density on the right as a function of the 
equilibrium density on the left. This implies that with this sequence of inequalities, the braid production and 
separation may be considered to take place in a time-averaged, effective potential where the 
production part is set by the 
compression forces and braids separate under a force related to extension. Specifically, we consider
\begin{equation}
    U(x) = 
    \begin{cases}
    A_c x , x < l 
    \\
    A_c l + B_s (x - l) , l < x < L
    \\
    \infty , x > L.
    \end{cases}
\end{equation}
In this potential, the probability of braids on the left $x<l$ will be proportional to 
\begin{equation}
    Z_1 = \frac{1}{\beta A_c} (1  -   e^{-\beta A_c l} ),
\end{equation}
while on the right
\begin{equation}
    Z_2 =  \frac{1}{\beta B_s}  e^{ -\beta (A_c - B_s) l}   ( e^{ -\beta B_s l } - e^{ -\beta B_s L }  ).
\end{equation}
The ratio is
\begin{equation}
    Z_2/ Z_1 \approx -\frac{A_c}{ B_s} e^{ -\beta (A_c - B_s) l}   e^{ -\beta B_s L } \gg 1
\end{equation}
which exceeds the case of only compression by a factor of $e^{ \beta | B_s | L }$, and the case of 
only expansion by a factor of $e^{\beta (A_s - A_c ) l }$ -- see Eq.~\ref{eq:equilibrium-ratio}.

\section{Constant force stretching dynamics of a braided bundle}
\label{sec:dynamic}

We now consider an experiment in which one stretches a previously 
compressed bundle (by laser tweezers or other means) at a fixed
force and determines the time rate of change of the bundle's length.  This is akin to a step 
force rheological measurement, and is closely related to determining the force-extension curve of a 
filament or filament bundle.  Typically, in such 
force extension measurements, one considers the limit of slow extension so that the 
observed length corresponds the thermal equilibrium prediction under a fixed force~\cite{Janmey1995}.  
In this case, however the extension of the bundle will be time dependent even though the force is constant.  

To study this problem, we minimize the energy of the bundle under a fixed stretching force.  
By doing so, we assume that bending of the kink angles at the braids is fast compared to the time 
scale of measurement of the end-to-end distance. We do not, however, assume that the advection and 
diffusion of the braids is similarly fast. Doing this energy minimization 
numerically, we obtain the time dependence shown in Fig.~\ref{fig:diffusion}. The time dependence of the bundle's 
extension is nonmonotonic: as defects diffuse from each other, the bundle initially becomes shorter. This happens 
due to the fact that as $x$ grows, the $L-x$ becomes smaller, hence, the moment of the force decreases as well. 
Since the stretching force decreases, kinks become less stretched and their angles increase, decreasing end-to-end distance. 

\section{Summary}
\label{sec:summary}
Semiflexible filament bundles admit three classes of topological defects: loops, braids, and dislocations.  These 
defects are all likely to be produced in the formation of bundles from solutions of semiflexible filaments by the
introduction of cross-linking agents.  They are, however, unlikely to form spontaneously in thermal 
equilibrium when the filament bundles are also chemically equilibrated with a reservoir of cross linkers. Here
we have pointed out that reciprocal compression and extension of filament bundles, however, is capable of 
producing a higher nonequilibrium density of braid defects within a bundle.  The key insight is that bundles 
under compression can locally buckle producing braid, anti-braid pairs. Upon subsequent extension, these braid
pairs will be driven to separate, as long as they do not immediately annihilate.  In order to assure that 
braid pairs produced in the previous compression cycle (and separated during the previous extension part of that
cycle) do not annihilate during the subsequent compressive cycle, one needs to introduce an asymmetry between 
the period of compression (short) and the period of extension (long).   Other cycles of compression and 
extension will also produce braid, anti-braid pairs, but at lower density.  We predict that the short 
compression period following long extension one will result in the maximum possible defect density. 

We also examined the extensional dynamics under a fixed tensile load 
of semiflexible bundles containing a braid defect pair. Here one
does not observe the standard worm-like chain force extension relation at very low frequencies.  The extension 
of the bundle is not controlled by the depletion of the length reservoir associated with the thermally-generated
undulations of the bundle, but rather by the motion of the braid defects and the bending of the kinks 
associated with these defects.  In essence, our predictions refer to the analog of plastic deformation in 
solids associated with defect motion rather than the (entropic) elastic response of the bundle, 
which, due to cross linking, is suppressed.  We note that the end-to-end distance of the bundle varies 
nonmonotonically with time under a constant tensile load. This somewhat counter-intuitive result occurs due to 
the combination of two effects: braid separation, which lengthens the bundle, and kink angle relaxation, which 
shortens it. 

For experimental verification of these predictions, there is no more direct measurement than
compression/extension experiments on individual semiflexible filament bundles using laser or magnetic tweezers 
to manipulate the bundle's stress state~\cite{Strehle2017}. We are currently unaware of such studies, but look forward to 
seeing them.  Although less direct in testing the predictions made here, standard shear measurements and studies
of stress relaxation in networks of filament bundles with transient cross linkers 
at long times or low frequencies are particularly relevant 
to the present work.  We imagine that, at sufficiently long times, stress relaxation will be dominated by 
plastic deformation of network comprised of both tearing and reattachment of bundles from each other, and the 
plastic deformation of the individual bundles themselves, presumably following the mechanisms discussed here.  
We do not, as yet, understand how to distinguish these dynamics in rheological data, and this remains one of 
the principal open questions related to this work.

\section*{Acknowledgements}
The authors acknowledge partial support from grant no.~NSF-DMR--1709785. V.M.S. also 
acknowledges support from the Bhaumik Institute Graduate Fellowship, and UCLA Dissertation Year Fellowship.

\section*{Author Contributions}
Conceptualization, Alex Levine; Formal analysis, Valentin Slepukhin; Funding acquisition, Alex Levine; Investigation, Valentin Slepukhin; Methodology, Alex Levine; Project administration, Alex Levine; Software, Valentin Slepukhin; Supervision, Alex Levine; Visualization, Valentin Slepukhin; Writing – original draft, Valentin Slepukhin and Alex Levine; Writing – review \& editing, Valentin Slepukhin and Alex Levine.

%%%END OF MAIN TEXT%%%


\balance


%%%REFERENCES%%%
\bibliography{thesis_references} %You need to replace "rsc" on this line with the name of your .bib file

\providecommand*{\mcitethebibliography}{\thebibliography}
\csname @ifundefined\endcsname{endmcitethebibliography}
{\let\endmcitethebibliography\endthebibliography}{}
\begin{mcitethebibliography}{8}
\providecommand*{\natexlab}[1]{#1}
\providecommand*{\mciteSetBstSublistMode}[1]{}
\providecommand*{\mciteSetBstMaxWidthForm}[2]{}
\providecommand*{\mciteBstWouldAddEndPuncttrue}
  {\def\EndOfBibitem{\unskip.}}
\providecommand*{\mciteBstWouldAddEndPunctfalse}
  {\let\EndOfBibitem\relax}
\providecommand*{\mciteSetBstMidEndSepPunct}[3]{}
\providecommand*{\mciteSetBstSublistLabelBeginEnd}[3]{}
\providecommand*{\EndOfBibitem}{}
\mciteSetBstSublistMode{f}
\mciteSetBstMaxWidthForm{subitem}
{(\emph{\alph{mcitesubitemcount}})}
\mciteSetBstSublistLabelBeginEnd{\mcitemaxwidthsubitemform\space}
{\relax}{\relax}

\bibitem[Heussinger and Frey(2007)]{Heussinger2007}
C.~Heussinger and E.~Frey, \emph{The European Physical Journal E}, 2007,
  \textbf{24}, 47--53\relax
\mciteBstWouldAddEndPuncttrue
\mciteSetBstMidEndSepPunct{\mcitedefaultmidpunct}
{\mcitedefaultendpunct}{\mcitedefaultseppunct}\relax
\EndOfBibitem
\bibitem[Broedersz \emph{et~al.}(2010)Broedersz, Depken, Yao, Pollak, Weitz,
  and MacKintosh]{Weitz2010}
C.~P. Broedersz, M.~Depken, N.~Y. Yao, M.~R. Pollak, D.~A. Weitz and F.~C.
  MacKintosh, \emph{Phys. Rev. Lett.}, 2010, \textbf{105}, 238101\relax
\mciteBstWouldAddEndPuncttrue
\mciteSetBstMidEndSepPunct{\mcitedefaultmidpunct}
{\mcitedefaultendpunct}{\mcitedefaultseppunct}\relax
\EndOfBibitem
\bibitem[M\"{u}ller \emph{et~al.}(2014)M\"{u}ller, Bruinsma, Lieleg, Bausch,
  Wall, and Levine]{Mueller2014}
K.~W. M\"{u}ller, R.~F. Bruinsma, O.~Lieleg, A.~R. Bausch, W.~A. Wall and A.~J.
  Levine, \emph{Phys. Rev. Lett.}, 2014, \textbf{112}, 238102--\relax
\mciteBstWouldAddEndPuncttrue
\mciteSetBstMidEndSepPunct{\mcitedefaultmidpunct}
{\mcitedefaultendpunct}{\mcitedefaultseppunct}\relax
\EndOfBibitem
\bibitem[Slepukhin \emph{et~al.}(2021)Slepukhin, Grill, Hu, Botvinick, Wall,
  and Levine]{slepukhin2021topological}
V.~M. Slepukhin, M.~J. Grill, Q.~Hu, E.~L. Botvinick, W.~A. Wall and A.~J.
  Levine, \emph{Proceedings of the National Academy of Sciences}, 2021,
  \textbf{118}, year\relax
\mciteBstWouldAddEndPuncttrue
\mciteSetBstMidEndSepPunct{\mcitedefaultmidpunct}
{\mcitedefaultendpunct}{\mcitedefaultseppunct}\relax
\EndOfBibitem
\bibitem[Slepukhin and Levine(2021)]{slepukhin2021thermal}
V.~M. Slepukhin and A.~J. Levine, \emph{arXiv preprint arXiv:2103.08832},
  2021\relax
\mciteBstWouldAddEndPuncttrue
\mciteSetBstMidEndSepPunct{\mcitedefaultmidpunct}
{\mcitedefaultendpunct}{\mcitedefaultseppunct}\relax
\EndOfBibitem
\bibitem[Kramers(1940)]{kramers1940brownian}
H.~A. Kramers, \emph{Physica}, 1940, \textbf{7}, 284--304\relax
\mciteBstWouldAddEndPuncttrue
\mciteSetBstMidEndSepPunct{\mcitedefaultmidpunct}
{\mcitedefaultendpunct}{\mcitedefaultseppunct}\relax
\EndOfBibitem
\bibitem[MacKintosh \emph{et~al.}(1995)MacKintosh, K\"as, and
  Janmey]{Janmey1995}
F.~C. MacKintosh, J.~K\"as and P.~A. Janmey, \emph{Phys. Rev. Lett.}, 1995,
  \textbf{75}, 4425--4428\relax
\mciteBstWouldAddEndPuncttrue
\mciteSetBstMidEndSepPunct{\mcitedefaultmidpunct}
{\mcitedefaultendpunct}{\mcitedefaultseppunct}\relax
\EndOfBibitem
\bibitem[Strehle \emph{et~al.}(2017)Strehle, Mollenkopf, Glaser, Golde,
  Schuldt, Käs, and Schnauß]{Strehle2017}
D.~Strehle, P.~Mollenkopf, M.~Glaser, T.~Golde, C.~Schuldt, J.~A. Käs and
  J.~Schnauß, \emph{Molecules}, 2017, \textbf{22}, year\relax
\mciteBstWouldAddEndPuncttrue
\mciteSetBstMidEndSepPunct{\mcitedefaultmidpunct}
{\mcitedefaultendpunct}{\mcitedefaultseppunct}\relax
\EndOfBibitem
\end{mcitethebibliography}
\bibliographystyle{rsc} %the RSC's .bst file

\end{document}